\documentstyle[11pt]{article}

\begin{document}


\title{A Quantum Structure Description of the Liar Paradox\footnote{Published as: Aerts, D.,
Broekaert, J. and Smets, S., 1999, ``A Quantum Structure Description of the Liar Paradox", {\it
International Journal of Theoretical Physics}, {\bf 38}, 3231-3239.}}
\author{Diederik Aerts,
Jan Broekaert and Sonja Smets}
\date{}
\maketitle
\centerline{Center Leo Apostel (CLEA),}
\centerline{Brussels Free University,}
\centerline{Krijgskundestraat 33, 1160 Brussels}
\centerline{diraerts@vub.ac.be, jbroekae@vub.ac.be}
\centerline{sonsmets@vub.ac.be}


\begin{abstract}
\noindent
In this article we propose an approach that models the truth behavior of
cognitive entities (i.e. sets
of connected propositions) by taking into account in a very explicit way
the possible influence of the
cognitive person (the one that interacts with the considered cognitive
entity). Hereby we specifically
apply the mathematical formalism of quantum mechanics because of the fact
that this formalism allows the
description of  real contextual influences, i.e. the influence  of the
measuring apparatus
on the physical entity. We concentrated on the typical situation of the
liar paradox and have shown that
(1) the truth-false state of this liar paradox can be represented by a
quantum vector of the non-product type in a finite
dimensional complex Hilbert space and the different cognitive
interactions by the actions of the
corresponding quantum projections, (2) the typical  oscillations
between false and truth - the paradox -is now quantum dynamically
described by a  Schr\"odinger equation.
We analyse possible philosophical implications of this result.
\end{abstract}

\section{Introduction.}

The liar paradox is the oldest semantical paradox we find in
literature.  In its simplest forms we trace
the paradox back to Eubulides - a pupil of Euclid - and to the
Cretan Epimenides. From the Greeks on
till today, different alternative forms of the liar emerged. We  now
encounter variations of the one
sentence paradox (the simplest form of the liar) but also of the two
or more sentence paradox. The two
sentence paradox is known as the postcard paradox of Jourdain, which
goes back to Buridan in 1300. On
one side of a postcard we read `the  sentence on the other side of
this card is true' and on the other
side of it we read `the sentence  on the other side of this card is
false'.

In this paper we will not work with the original forms of the
paradox,  but in the version in which we
use an index or sentence pointer followed by  the sentence this index
points at :

\begin{center}
\underline{\sl  Single Liar :}

\medskip

(1)  \ \ \    sentence (1) is false

\bigskip
\underline{\sl Double Liar :}

\medskip
 (1)   \ \ \    sentence (2) is false

\medskip
 (2)    \ \ \   sentence (1) is true

\end{center}

\section{Applying the Quantum Mechanical Formalism.}

The theories of chaos and complexity have shown that similar patterns of
behaviour can be
found in very different layers of reality.  The success of these theories
demonstrates that interesting conclusions about the nature of reality can
be inferred from the
encountered structural similarities  of
dynamical behaviour in different regions of reality. Chaos and
complexity theories are  however
deterministic theories that do not take into account the fundamental
contextuality that  is introduced by
the influence of the act of observation on the observed. Most of the
regions of reality are highly
contextual (e.g. the social layer, the cognitive layer, the  pre-
material quantum layer), rather with
exception of the material layer of reality were  contextuality is
minimal. In this sense it is strange
that no attempts have been undertaken to find  similarities using
contextual theories, such as quantum
mechanics, in the different  regions of reality. The study that we
present in this paper should be
classified as such an  attempt, and is part of one of the projects in
our center focusing on the layered
structure of reality (Clea Research Project,1997-; Aerts, 1994; Aerts, 1999)

We justify the use of the mathematical formalism of quantum mechanics to
model context dependent
entities, because a similar approach has already been developed by some
of us for the situation of an
opinion pole within the social layer of reality (Aerts, 1998; Aerts and
Aerts, 1994, 1997; Aerts, Broekaert and Smets, 1999; Aerts, Coecke and Smets, 1999). In such an opinion pole specific 
questions are put forward that introduce a real influence of the interviewer on the
interviewee, such that the situation is contextual. It is shown explicitly in (Aerts
and Aerts, 1995, 1997) that the probability model
that results is this situation is of a quantum mechanical nature.

By means of a model we will present the liar - one sentence - or the
double  liar - a group of sentences -  as one entity that we
consider to `exist' within the cognitive layer of reality. The
existence is being expressed by the
possibility of influencing other cognitive entities, and by the
different states that it can be in.
Indeed it has been shown that the concept of entity can be introduced
rigorously and founded on the
previously mentioned properties. In this way we justify the present
use (Aerts, 1992).

\section{Measuring Coginitive Entities : Modeling Truth Behavior.}

In this paragraph we will explore the context dependence of cognitive
entities like the liar paradox. We
introduce the explicit dependence of the truth
and falsehood of a sentence on the cognitive interaction with the
cognitive person. Reading a sence, or with other words `making a sentence
true or false' will be modeled as `performing a  measurement' on
the sentence within the cognitive
layer of reality. This means that in our  description a sentence
within the cognitive layer of reality is
`in general' neither true, nor  false. The `state true' and the
`state false' of the sentence are
`eigenstates' of the  measurement. During the act of measurement the
state of the  sentence changes in
such a way that it is true or that it is false. This general `neither
true nor false  state' will be
called a superposition state in analogy with the quantum  mechanical
concept. We shall see that it is
effectively a superposition state in the mathematical  sense after we
have introduced the complex Hilbert
space description.

We proceed operationally as follows. Before the cognitive measurement
(this means before we start to
interact with the sentence, read it and make a hypothesis about its
truth or falsehood) the sentence is
considered to be neither true nor false and  hence in a superposition
state. If we want to start to
analyse the cognitive inferences entailed, we make one of the two
possible hypothesis, that it is  true or that it is false. The making
of one of these two hypothesis -
this is part of the  act of measurement - changes the state of the
sentence to one of the two eigenstates
- true or  false. As a consequence of the act of measurement the
sentence becomes true or false  (is in
the state true or false) within the cognitive entity were the
sentence is part of. This change
influences the state of this complete cognitive entity. We will see
that if  we apply this approach to the
double liar, that the change of state puts into work a dynamic
process that we can describe by a
Schr\"odinger equation.
We  have to consider three situations:
\[
{\rm A}\ \   \left\{
\begin{array} {ll}
 {\rm (1)\  }   &  {\rm sentence\ (2)\ is\ false} \\ {\rm (2)\ }   &
{\rm sentence\ (1)\ is\ true}
\end{array} \right.
\]

\[
{\rm B}\ \   \left\{
\begin{array} {ll}
 {\rm (1)\  }   &  {\rm sentence\ (2)\ is\ true} \\ {\rm (2)\ }   &  {\rm
sentence\ (1)\ is\ true}
\end{array} \right.
\]

\[
{\rm C} \ \   \left\{
\begin{array} {ll}
 {\rm (1)\  }   &  {\rm sentence\ (2)\ is\ false} \\ {\rm (2)\ }   &
{\rm sentence\ (1)\ is\ false}
\end{array} \right.
\]
\section{The Double Liar: A Full Quantum Description.}
 The resemblance of the truth values of single sentences and the two-fold
eigenvalues of a spin-1/2 state is used to construct a dynamical
representation; the measurement evolution as well as a continuous time
evolution are included.

We recall some elementary properties of a spin state. Elementary
particles - like the electron - are bestowed with a property referred to as an
intrinsic angular momentum or spin. The spin of a
particle is quantised: upon measurement the particle only exposes a finite
number of distinct spin values. For the spin-1/2 particle, the number of spin
states is two, they are commonly referred to as the `up' and `down' state.
This two-valuedness can adequately describe the truth function of a liar type
cognitive entity. Such a sentence supposedly is either true or false. The
quantum mechanical description on the other hand allows a superposition of the
`true' and `false' state. This corresponds to our view of allowing cognitive
entities before measurement - i.e. reading and hypothetising - to reside in a
non-determinate state of truth or falsehood. In quantum  mechanics such a
state $\Psi$ is described by a poundered superposition of the two states:
\[
 \Psi = c_{ true} \left( \begin{array}{c}  1 \\ 0 \end{array}
\right)  +
 c_{ false} \left(\begin{array}{c}  0 \\ 1 \end{array} \right)
\] The operation of finding whether such a cognitive entity is true or false,
is done by applying respectively the true-projector $P_{true}$ or
false-projector
$P_{false}$.
\[
 P_{true} = \left( \begin{array}{cc}  1 & 0 \\ 0 & 0
  \end{array} \right)  \ \ \ \ \ \  P_{false } = \left(
\begin{array}{cc}  0 & 0 \\ 0 & 1
  \end{array} \right)
\] In practice in the context of the cognitive entity, this corresponds to the
assignment of either truth or falsehood to a sentence after its reading. In
quantum mechanics, the true-measurement on the superposed state
$\Psi$ results in the true state ;
\[
 P_{true} \Psi = c_{true}\left( \begin{array}{c}  1 \\ 0
\end{array} \right)
\]
while the square modulus of the corresponding pounderation factor $c_{
true}$ gives the statistical probability of finding the entity in the
true-state. An unequivocal result is therefore not obtained
when the superposition does not leave out one of the states completely, i.e.
either $c_{true}$ or $c_{false}$  is zero. Only in those instances do we
attribute to a sentence its truth or falsehood.

The coupled sentences of
the two-sentence liar paradox (C) for instance are precisely described by the
so called `singlet state'. This global state combines, using the tensor
product $\otimes$, states of sentence one with states of sentence two:
\[
\frac{1}{\sqrt{2}}\left\{ \left( \begin{array}{c}  1  \\ 0
\end{array} \right)  \otimes \left(
\begin{array}{c}  0 \\ 1 \end{array} \right)  -  \left(
\begin{array}{c}  0  \\ 1 \end{array} \right)
\otimes \left( \begin{array}{c}  1 \\ 0 \end{array} \right)  \right\}
\] The appropriate true-projectors for sentence one and two are now:
\[
 P_{1,true} = \left( \begin{array}{cc}  1 & 0 \\ 0 & 0
  \end{array} \right) \otimes 1_2 \ \ \ \ \ \  P_{2,true } = 1_1
\otimes \left( \begin{array}{cc}  1 & 0
\\ 0 & 0
  \end{array} \right)
\] The false-projectors are obtained by switching the diagonal elements $1$
and $0$ on the diagonal of the matrix.

In the same manner the coupled
sentences of the liar paradox (B) can be constructed:
\[
\frac{1}{\sqrt{2}}\left\{ \left( \begin{array}{c}  1  \\ 0
\end{array} \right)  \otimes \left(
\begin{array}{c}  1 \\ 0
\end{array} \right)  -  \left( \begin{array}{c}  0  \\ 1 \end{array}
\right)  \otimes \left(
\begin{array}{c}  0 \\ 1 \end{array} \right)  \right\}
\] Our final aim is to describe the real double liar paradox (A) quantum
mechanically and even more to show how the true-false cycle originates from
the Schr{\"o}dinger time-evolution of the appropriate initial state.  The
description of this system necessitates the coupled Hilbert space $C^4\otimes
C^4$, a larger space than for the previous systems. In this case the truth
and falsehood values from measurement and semantical origin must be discerned,
the dimension for each sentence therefore must be 4.

The initial unmeasured state - i.e. $\Psi_0$ -  of the real double liar
paradox is: {\small\[
\frac{1}{2}\left\{ \left( \begin{array}{c} 0\\ 0\\ 1 \\ 0 \end{array}
\right)  \otimes \left(
\begin{array}{c}  0 \\ 1 \\ 0 \\ 0 \end{array} \right)  +
 \left( \begin{array}{c} 0\\ 1\\ 0 \\ 0 \end{array} \right)  \otimes
\left( \begin{array}{c}  0 \\ 0 \\ 0
\\ 1 \end{array} \right)  +
\left( \begin{array}{c} 0\\ 0\\ 0 \\ 1 \end{array} \right)  \otimes
\left( \begin{array}{c} 1 \\ 0 \\ 0
\\ 0
\end{array} \right) +
\left( \begin{array}{c} 1\\ 0\\ 0 \\ 0 \end{array} \right)  \otimes
\left( \begin{array}{c}  0 \\ 0 \\ 1
\\ 0
\end{array} \right) \right\}
\]} Each next term in this sum is actually the consecutive state which is
reached in the course of time, when the paradox is read through. This can be
easily verified by applying the appropriate truth-operators:
\[
 P_{1,true} = \left( \begin{array}{cccc}  0 & 0 & 0 & 0 \\ 0 & 0 & 0 & 0 \\ 0
& 0 & 1 & 0 \\0 & 0 & 0 & 0
  \end{array} \right) \otimes 1_2 \ \ \ \ \ \
 P_{2,true } = 1_1 \otimes\left( \begin{array}{cccc}  0 & 0 & 0 & 0
\\ 0 & 0 & 0 & 0 \\ 0 & 0 & 1 & 0 \\ 0 & 0 & 0 & 0 \end{array} \right)
\] The projectors for the false-states are constructed by placing the
$1$ on the final diagonal place.

The explicit construction of the unitary evolution operator is accomplished
through an intermediary equivalent representation in $C^{16}$. The complex
space
$C^4\otimes C^4$ is isomorphic to $C^{16}$.
In this aim the basis of the $C^{16}$ is constructed as (  $i$ and $j$ from
1 to 4 )  :
\[
e_i  \otimes e_j = e_{ \kappa(i,j) } \ \ \ {\rm and} \ \ \ \kappa (i,j) =
4(i-1) +j
\]
In $C^{16}$ the unmeasured state $\Psi_0$ is then given by:
\[
\Psi_0 = \frac{1}{2} \{ e_{10} + e_{8} + e_{13} + e_{3} \}
\]
The 4 by 4 submatrix -   $U_D$ -  of the discrete unitary evolution operator,
which describes the time-evolution at instants of time when a sentence has
changed truth value, is:
\[  U_D =
\left( \begin{array}{cccc}   0 & 0 & 0 & 1 \\   0 & 0 & 1 & 0 \\ 1 & 0 & 0 & 0
\\ 0 & 1 & 0 & 0
\end{array}\right)
\]
In order to obtain a description at every instance of time, a procedure
of diagonalisation on the submatrix $U_D$ was performed, i.e. $ U_D |_{\rm
diag}$. From the Schr\"odinger evolution and Stone's Theorem we obtain:
\[
 H_{sub} |_{\rm diag} =i \ln  U_D|_{\rm diag}
\]
Now inverting the procedure of diagonalisation, the infinitesimal
generator of the time-evolution - the submatrix hamiltonian - is obtained :
\[  H_{sub} =
\left( \begin{array}{cccc}
-1/2&-1/2&(1-i)/2&(1+i)/2 \\
-1/2&-1/2&(1+i)/2&(1-i)/2 \\
(1+i)/2&(1-i)/2&1/2&1/2 \\
(1-i)/2&(1+i)/2&1/2&1/2
\end{array}\right)
\]
The submatrix of the evolution operator $U(t)$, valid at all times is then
given by the expression:
\[  U_{sub}(t) = e^{- i H_{sub} t}
\]
The time evolution operator $U_{sub}(t)$ in the 4 by 4 subspace of $C^{16}$
becomes (modulo a numerical factor $\frac{1}{4}$ for all elements ):
{\small\[
\left( \begin{array}{llll}
1 + e^{-i t} + e^{i t} +e ^{2 i t} &
1 - e^{- i t} - e^{i t} + e^{2 i t} &
1 - i e^{- i t} + i e^{i t} -e^{2 i t} &
1 + i e^{- i t} - i e^{i t} -e ^{2 i t}
\\
1 - e^{-i t} - e^{i t} +e ^{2 i t} &
1 + e^{- i t} + e^{i t} + e^{2 i t} &
1 + i e^{- i t} - i e^{i t} -e^{2 i t} &
1 - i e^{- i t} + i e^{i t} -e ^{2 i t}
\\
1 + i e^{-i t} - i e^{i t} - e ^{2 i t} &
1 -i  e^{- i t} + i e^{i t} - e^{2 i t} &
1 + e^{- i t} +  e^{i t} + e^{2 i t} &
1 - e^{- i t} -  e^{i t} + e ^{2 i t}
\\
1 - i e^{-i t} + i e^{i t} - e ^{2 i t} &
1 +i  e^{- i t} - i e^{i t} - e^{2 i t} &
1 - e^{- i t} -  e^{i t} + e^{2 i t} &
1 + e^{- i t} +  e^{i t} + e ^{2 i t}
\end{array}\right)
\] }
The hamiltonian $H$ as well as the time-evolution operator $U(t)$ in
$C^4\otimes C^4$  is immediately obtained by inverting
the basis transformation function $\kappa$:
\[
 H =\sum_{\kappa, \lambda = 1}^{16} {H _{sub}}_{\kappa(i,j) \lambda(u,v)}
O_{i u}\otimes O_{j v}
\]
and
\[
 U(t) =\sum_{\kappa, \lambda = 1}^{16} {U _{sub}}_{\kappa(i,j) \lambda(u,v)}(t)
O_{i u}\otimes O_{j v}
\]
with;
\[
O_{i u}\otimes O_{j v}=\{ e_i.e_u^t \}\otimes\{ e_j.e_v^t \}
\]
For example, term $\kappa = 3$ , $\lambda  = 10$ of the time evolution
operator $U(t)$ is;
\[
\frac{1}{4}(1- i e^{-i t} +i e^{-i t} - i e^{ 2 i t})
\left( \begin{array}{cccc}   0 & 0 & 1 & 0 \\   0 & 0 & 0 & 0 \\ 0 & 0 & 0 & 0
\\ 0 & 0 & 0 & 0
\end{array}\right) \otimes \left( \begin{array}{cccc}   0 & 0 & 0 & 0 \\   0 &
0 & 0 & 0 \\ 0 & 1 & 0 & 0
\\ 0 & 0 & 0 & 0
\end{array}\right)
\]
Starting from the  initial state $\Psi_0$ the constructed dynamical evolution
leaves the system unchanged; $\Psi_0$ is a time invariant state:
\[
\Psi_0 (t) = \Psi_0
\]
 As soon as a measurement for truth or falsehood on either of the sentences
is made, the
dynamical evolution sets of in a cyclical mode, attributing alternatively
thruth and falsehood to the consecutively read sentences.

The quantum formalism therefore seems an appropriate tool to describe the
liar paradox.
Could the formalism be applied to more intricate cognitive entities?
Given the procedure we applied -  an adaptation of the formalism of two
interacting spin-3/2 particles - it is possible to extend the liar paradox
to more complex variants of multiple sentences refering to one another in a
truth confirming or denying manner. The minimal dimension to represent
quantummechanically such a paradoxical set of $n$ sentences will not be less
than $2^n$. The exact dimension of the appropriate Hilbert space depends on
the
specific n-sentence liar paradox described.

\section{Conclusion.}

We analysed how cognitive entities behave by using the formalism of
quantum mechanics where the influence of the cognitive observer on the
cognitive  entity can be taken into
account. In the same way as we described the double liar we can
also represent the n-dimensional liar. The vector in the Hilbert space that
we used to  represent the
state of the double liar is an
eigenvector of the Hamiltonian of the system. This shows that we can
consider the double liar as a
cognitive entity without being measured on as an invariant of the
time evolution. Once a measurement - a
cognitive act - on one of the  sub-elements is performed, the whole
cognitive entity changes into a state
that is no eigenstate anymore of the Hamiltonian. And after this
measurement this state will start to
change dynamically in the typical way of the liar paradox, sentences
becoming
true and false, and staying constantly
coupled. This  behaviour is exactly described by the Schr\"odinger
equation that we have derived. In
this way we have given a description of the internal dynamics within
self-referring cognitive
entities as the liar paradox. Our aim is to develop this approach
further and to analyse in which
way we can describe other examples of cognitive entities. We also want to
analyse in further research in
which way this result can be interpreted within a general scheme that
connects different layers of reality
structurally. Some profound philosophical questions, still very
speculative at this stage of our
research, but certainly stimulating, can be put forward: e.g. Can we
learn something about the nature
and origin of dynamical change by considering this example of the liar
paradox? Could the cognitive
layer be considered being in a very early structuring stage, such that we
trace down very primitive
dynamical and contextual processes that could throw some light on
primitive dynamical and contextual
processes encountered in the pre-material layer (e.g. spin processes)?
Apart from these speculative but
stimulating philosophical questions, we also would like to investigate
further in which way our quantum
mechanical model for the cognitive layer of reality could be an
inspiration for the development of a
general interactive logic that can take into acount
more subtle dynamical
and contextual influences
than just those of the cognitive person on the truth behavior of the
cognitive entities.

\section{Acknowledgements}

Diederik Aerts is Senior Research Associate of the Fund
for Scientific Research, Flanders, Jan Broekaert is Post Doctoral researcher at the
AWI-grant Caw96/54a of the Flemish Community and Sonja Smets is Research Assistant of
the Fund for Scientific Research Flanders. Part of the research presented in this
article was realised with the aid of AWI-grant Caw96/54a of the Flemish Community.

\section{References.}

\begin{description}

\item Aerts, D., 1992, ``Construction of reality and
its influence on the understanding of quantum structures'', {\it  Int. J.
Theor. Phys.}, {\bf 31}, 1813.

\item
Aerts, D., 1994, ``The Biomousa: a new view of discovery and creation", in
{\it Perspectives on the World, an interdisciplinary reflection}, eds. Aerts, D., et al., VUBPress.

\item
Aerts, D., 1998, ``The entity and modern physics: the creation-discovery view of reality", in {\it Interpreting Bodies:
Classical and
      Quantum Objects in Modern Physics}, ed. Castellani, E., Princeton University Press, Princeton.

\item 
Aerts, D., 1999, ``The game of the biomousa: a view of discovery and creation, in {\it Worldviews and the problem of
synthesis}, eds. Aerts, D., Van Belle, H. and Van der Veken, J., KLuwer Academic, Dordecht.

\item
Aerts, D. and Aerts, S., 1994, ``Applications of quantum statistics
in psychological studies of decision processes", {\it Foundations of
Science} {\bf 1}, 85.

\item
Aerts, D. and Aerts, S., 1997, ``Applications of quantum statistics
in psychological studies of decision processes", in {\it Foundations of Statistics},
eds. Van Fraassen B., Kluwer Academic, Dordrecht.

\item
Aerts, D. Broekaert, J. and Smets, S., 1999, ``The liar paradox in a
quantum mechanical perspective", {\it Foundations of Science}, {\bf 4}, 115.

\item
Aerts, D. Coecke B. and Smets S., 1999, ``On the origin of probabilities
in quantum mechanics: creative and contextual aspects'', in {\it Metadebates On Science},
eds. Cornelis, G., Smets, S. and Van Bendegem, J.P., Kluwer Academic, Dordrecht.

\item
Clea Research Project (1997-2000), ``Integrating Worldviews: Research on the
Interdisciplinary Construction
of a Model of Reality with Ethical and Practial Relevance'' Ministry of
the Flemish Community, dept.
Science, Innovation and Media.

\end{description}

\end{document}